\documentclass[aps,prl,twocolumn,floatfix,bibliography,superscriptaddress]{revtex4}
\usepackage{graphics,bm,amsmath,amssymb,natbib,url,epsfig,psfrag,xcolor}
\usepackage{graphicx}
\usepackage{subfigure}
\usepackage[normalem]{ulem}
\usepackage[mono]{inconsolata}
\usepackage{xspace}
\def\QSpace{\texttt{QSpace}\xspace} 

\usepackage[unicode=true,pdfusetitle,
bookmarks=true,bookmarksnumbered=false,bookmarksopen=false,
breaklinks=false,pdfborder={0 0 0},pdfborderstyle={},backref=false,colorlinks=true]
{hyperref}

\hypersetup{colorlinks,allcolors=blue}

\begin{document}
	\newcommand{\Tr}{\text{Tr}}
	\newcommand{\beq}{\begin{equation}}
		\newcommand{\eeq}{\end{equation}}
	\newcommand{\eran}[1]{{\color{blue}#1}}
	
	\newcommand{\be}{\begin{equation}}
		\newcommand{\ee}{\end{equation}}
	\newcommand{\bea}{\begin{eqnarray}}
		\newcommand{\eea}{\end{eqnarray}}
	\newcommand{\ES}[1]{{\color{violet}#1}}
	\newcommand{\sarath}[1]{{\color{blue}#1}}
	\newcommand{\YM}[1]{{\color{red}#1}}
	
	\def\bs#1\es{\begin{split}#1\end{split}}	\def\bal#1\eal{\begin{align}#1\end{align}}
	\newcommand{\nn}{\nonumber}
	\newcommand{\sgn}{\text{sgn}}
	
	\title{Back-action effects in charge detection }

	\author{Sarath Sankar}
	\email[]{sarathsankar@tauex.tau.ac.il}
	\affiliation{School of Physics and Astronomy, Tel Aviv University, Tel Aviv 6997801, Israel}
	\author{Matan Lotem}
	\affiliation{Department of Physics, Ben-Gurion University of the Negev, Beer-Sheva 84105, Israel}
	\affiliation{Department of Physics, Princeton University, Princeton, NJ 08544, USA}
	\author{Joshua Folk}
	\affiliation{Quantum Matter Institute, University of British Columbia, Vancouver, British Columbia, Canada}
	\affiliation{Department of Physics and Astronomy, University of British Columbia, Vancouver, British
		Columbia, Canada}
	\author{Eran Sela}
	\affiliation{School of Physics and Astronomy, Tel Aviv University, Tel Aviv 6997801, Israel}
	\author{Yigal Meir}
	\affiliation{Department of Physics, Ben-Gurion University of the Negev, Beer-Sheva 84105, Israel}

	\date{\today}
	\begin{abstract}
		Charge detection offers a powerful probe of mesoscopic structures based on quantum dots, but it also invariably results in measurement back-action (MBA).  If strong, MBA can be detrimental to the physical properties being probed. In this work, we focus on the effects of MBA on an Anderson impurity model in which the impurity is coupled electrostatically to a detector. Introducing a novel non-perturbative method, we explore the interplay of coherent dynamics, strong correlations and non-equilibrium conditions. The effects of MBA can be seen most clearly in the temperature derivative of occupation. In the non-equilibrium case, we identify this as arising due to an energy flow from the detector to the impurity.
	\end{abstract}
	\maketitle
	\paragraph{Introduction.}
	Charge  measurement is a valuable experimental probe in mesoscopic systems.
	The standard technique \cite{field1993measurements} involves a nearby detector, usually a quantum point contact or a quantum dot, the current through which is sensitive to the charge of the system due to electrostatic interactions. Such charge detectors have been widely used to study a variety of physical phenomena such as Coulomb blockade \cite{field1993measurements}, dephasing in electron interference \cite{buks1998dephasing}, double-quantum-dot qubits \cite{elzerman2003few,petta2004manipulation}, full counting statistics \cite{gustavsson2006counting}, charge Kondo screening \cite{piquard2023observing}  and non-equilibrium thermodynamics \cite{saira2012test,koski2013distribution,hofmann2016equilibrium,barker2022experimental}. A 
	recent addition to the above list is the measurement of thermodynamic entropy using a Maxwell relation (MR) \cite{hartman2018direct,child2022entropy,adam2024entropy}, which has several promising applications ~\cite{cooper2009observable,yang2009thermopower,viola2012thermoelectric,ben2013detecting,ben2015detecting,smirnov2015majorana,sela2019detecting,hou2012ettingshausen,han2022fractional,sankar2023measuring}.

	While such charge measurements may often be considered as ``noninvasive" \cite{field1993measurements}, it has now been established that there are circumstances where the detector may have significant effects on the mesoscopic system via measurement back-action (MBA)\cite{buks1998dephasing,aleiner1997dephasing,gurvitz1997measurements,levinson1997dephasing,avinun2004controlled, kung2009noise, snyman2007polarization, young2010inelastic, zilberberg2014measuring, bischoff2015measurement,ferguson2023measurement, ma2023identifying, sankar2024detector}. 
	The charge detector continuously measures the system and 
	tends to suppress its unitary dynamics~\cite{szyniszewski2019entanglement,szyniszewski2020universality,kells2023topological,ma2023identifying}, as in the quantum Zeno effect~\cite{itano1990quantum,facchi2001quantum,koshino2005quantum}. In addition, a voltage-biased detector can also serve as an energy bath, leading to detector-assisted dynamics ~\cite{snyman2007polarization,young2010inelastic,zilberberg2014measuring,bischoff2015measurement,ferguson2023measurement}.

	The presence of the detector leads to a competition between strong correlations developing in the system itself (the Kondo effect in the example treated below), and the correlations and entanglement between the system and the detector, manifested, e.g., in the Anderson overlap catastrophe~\cite{matveev1992interaction,aleiner1997dephasing}. Here we look in detail at a simple example, a quantum dot (QD) coupled to a lead, and study how MBA due to the charge detector modifies its thermodynamic observables. We focus on the average charge of the QD, $N$, and its temperature derivative, $dN/dT$. The latter has escaped attention in previous MBA studies but encodes important thermodynamic information of the system. We show that $dN/dT$ is particularly sensitive to MBA. This makes it a robust marker for experimental investigation of back-action effects.
	
	This problem serves as a paradigmatic example of the rich interplay between strong correlations, coherent dynamics, and out-of-equilibrium physics, which  cannot at present be treated exactly using analytical or numerical methods.  Our treatment of the MBA is non-perturbative in both the QD-lead tunneling and the system-detector interaction. The latter is necessary to capture the many-body effect of the Anderson orthogonality catastrophe. 
	In contrast, previous MBA treatments are perturbative in the system-detector interaction. This includes the phenomenological $P(E)$ theory~\cite{ingold1992charge,aguado2000double,gustavsson2007frequency}, which describes how fluctuations in the electric field due to the current noise in the detector~\cite{kouwenhoven1994observation,platero2004photon} lead to a dephasing effect in the measured QD.

	\begin{figure}[t]
		\includegraphics[width=\columnwidth]{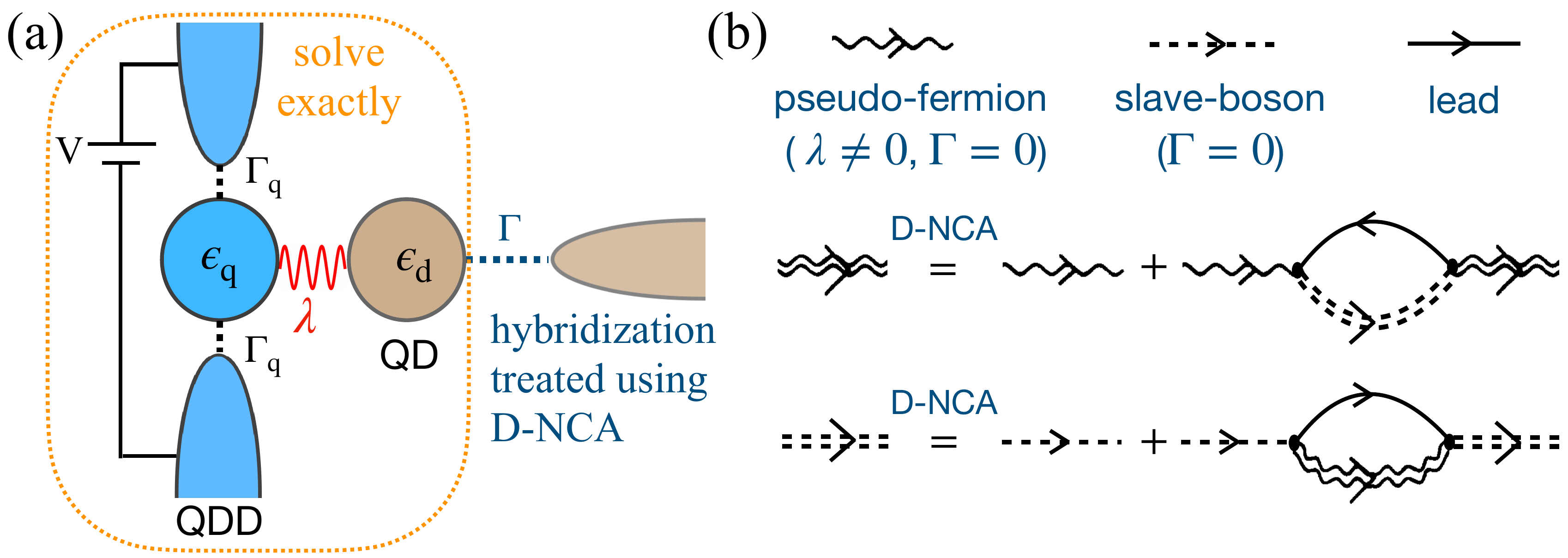}
		\caption{\label{fig:setup}(a) Schematic of an Anderson impurity model electrostatically coupled to a quantum dot detector (QDD). (b) For $\Gamma\to 0$, we calculate  the exact QD propagator including the interaction with the QDD (wiggly line), and then use it to carry out an hybridization expansion in $\Gamma$  by the D-NCA method discussed in the main text.
		} 
	\end{figure}
	
	\paragraph{Model.}
	We consider a QD emulating an Anderson impurity,  capacitively coupled to a detector as shown in Fig.~\ref{fig:setup}(a). For concreteness, we take the detector to be a quantum-dot detector (QDD); with an eye to future experimental tests, this choice has the advantage that the strength of the MBA due to a QDD can readily be tuned by changing the QDD energy level and its coupling to the detector leads~\cite{sankar2024detector}. However, the treatment in this paper will be formulated in terms of universal MBA parameters that are not specific to the type of detector.  
	
	The Hamiltonian is made up of parts representing the system, the detector, and the interaction between them,
	$H = H_{\rm{sys}}+ H_{\rm{det}}+H_{\rm{int}}$. The system is described by
	\begin{equation}
		\begin{split}
			H_{\rm{sys}} &= \sum_{\sigma}\epsilon_d d_\sigma^\dagger d_\sigma + Ud_\uparrow^\dagger d_\uparrow d_\downarrow^\dagger d_\downarrow \\
			&+\sum_{k\sigma}\epsilon_k c^\dagger_{k\sigma}c_{k\sigma} + \sum_{k\sigma}\left(\gamma_k c^\dagger_{k\sigma}d_\sigma + {\rm{H. c.}}\right).
		\end{split}
	\end{equation}
	Here, $d^\dagger_\sigma$ ($c^\dagger_{k\sigma}$) creates an electron in the single-level QD (lead) with spin $\sigma$ and energy $\epsilon_d$ ($\epsilon_k$). For later convenience, the density of states (DoS) of the lead is taken to be a Lorentzian of width $D$, so the line width (hybridization) of the QD is $\Delta(\omega)=2\pi  \sum_k |\gamma_k|^2 \delta(\omega-\epsilon_k)=\Gamma \frac{D^2}{\omega^2+D^2}$. We assume $\epsilon_d, \Gamma, T \ll D$ so that band-edge effects are negligible. The average charge in the QD is $N= \sum_{\sigma} \langle d^\dagger_\sigma d_\sigma\rangle$. The onsite interaction $U$ is taken to be infinite such that double occupancy is prohibited.
	
	The detector consists of a quantum dot tunnel-coupled to two leads at bias voltage $\pm V/2$, and described by
	\begin{equation}
		H_{\rm{det}}= \sum_{i=1}^m \epsilon_{q}q^\dagger_i q_i +\sum_{ik\mu} \left[ \epsilon_k \varphi_{ki\mu}^\dagger\varphi_{ki\mu} + v_{\mu}(\varphi_{ki\mu}^\dagger q_i + {\rm H.c.}) \right].
	\end{equation}
	Here, $q_{i}^\dagger$ ($\varphi_{ki\mu}^\dagger$) creates an electron in the detector dot (lead $\mu=L,R$) with energy $\epsilon_q$ ($\epsilon_k$)  and $i$ denotes additional conserved degrees of freedom, which can include spin and other channels; for simplicity, we treat it as a degeneracy factor with $m$ flavors. 
	The width of the detector QD level due to tunneling is $\Gamma_q=2\pi \nu_q (v_L^2+v_R^2)$ where $v_\mu$ represents the tunneling strength and $\nu_q$ is the DoS of the detector leads. The interaction strength between the system and detector is parameterized by $\lambda$, and given by 
	\begin{equation}
		H_{\rm{int}}=\lambda \sum_{i=1}^m\left(q^\dagger_i q_i -\frac{1}{2}\right)\left(\sum_\sigma d^\dagger_\sigma d_\sigma -\frac{1}{2}\right). 
	\end{equation}
	The QDD model ignores intra-detector-dot interactions to elucidate the MBA effects in a simple and universal manner. A treatment of MBA effects due to an interacting detector will be considered in a future work.

	\paragraph{Key results.} Our main result is a condition for the regime in which MBA on thermodynamic quantities can be neglected, and the identification of MBA effects that occur away from this regime. The strength of MBA is parameterized by two quantities: 
	\begin{itemize}
		\item $\alpha$: a dimensionless exponent that characterizes the strength of the Anderson orthogonality catastrophe. It is given by $\alpha=m(\delta/\pi)^2$, in terms of the change in the detector scattering phase shift, $\delta=\arctan\left(\frac{\epsilon_q+\lambda/2 }{\Gamma_q}\right)-\arctan\left(\frac{\epsilon_q-\lambda/2 }{\Gamma_q}\right)$, due to an electron tunneling in/out of the QD \cite{sankar2024detector}.
		\item $\Gamma_\varphi$: a dephasing rate that captures the inelastic effects of MBA due to voltage bias in the detector. It was introduced previously ~\cite{aleiner1997dephasing,levinson1997dephasing} in the context of dephasing of Aharonov-Bohm oscillations and is reproduced in Eq.~(\ref{eq:6}).
	\end{itemize}
	As we demonstrate below, MBA on thermodynamic observables is negligible provided that $\alpha\ll 1$ and $\Gamma_\varphi\ll T$. 
	At equilibrium $(V=0)$, MBA is governed only by $\alpha$, which reduces the width of the charging curve and modifies the lineshape of $dN/dT$.
	
	Most of our calculations apply to the non-equilibrium case, $(V\ne 0)$, for which we develop a new non-perturbative approach to solve the Anderson model coupled to a detector, capturing large electrostatic coupling to the detector (resulting in a large $\alpha)$ and also hybridization $(\Gamma)$ to the lead. We apply this method to show that the charging curve broadens due to $\Gamma_\varphi$, though this effect is small when $\Gamma_\varphi\ll \mathrm{max}(\Gamma,T)$.  Then, we confirm that the Maxwell relation based on $dN/dT$ may be applied only when $\Gamma_\varphi\ll T$. 
	Finally, we use the techniques developed in this paper to study the heat current that flows from a biased detector to the system, uncovering a strong but unexpected dependence of the heat current on occupation.
	
	\paragraph{Equilibrium MBA effects.}
	
	\begin{figure}[t]
		\includegraphics[width=\columnwidth]{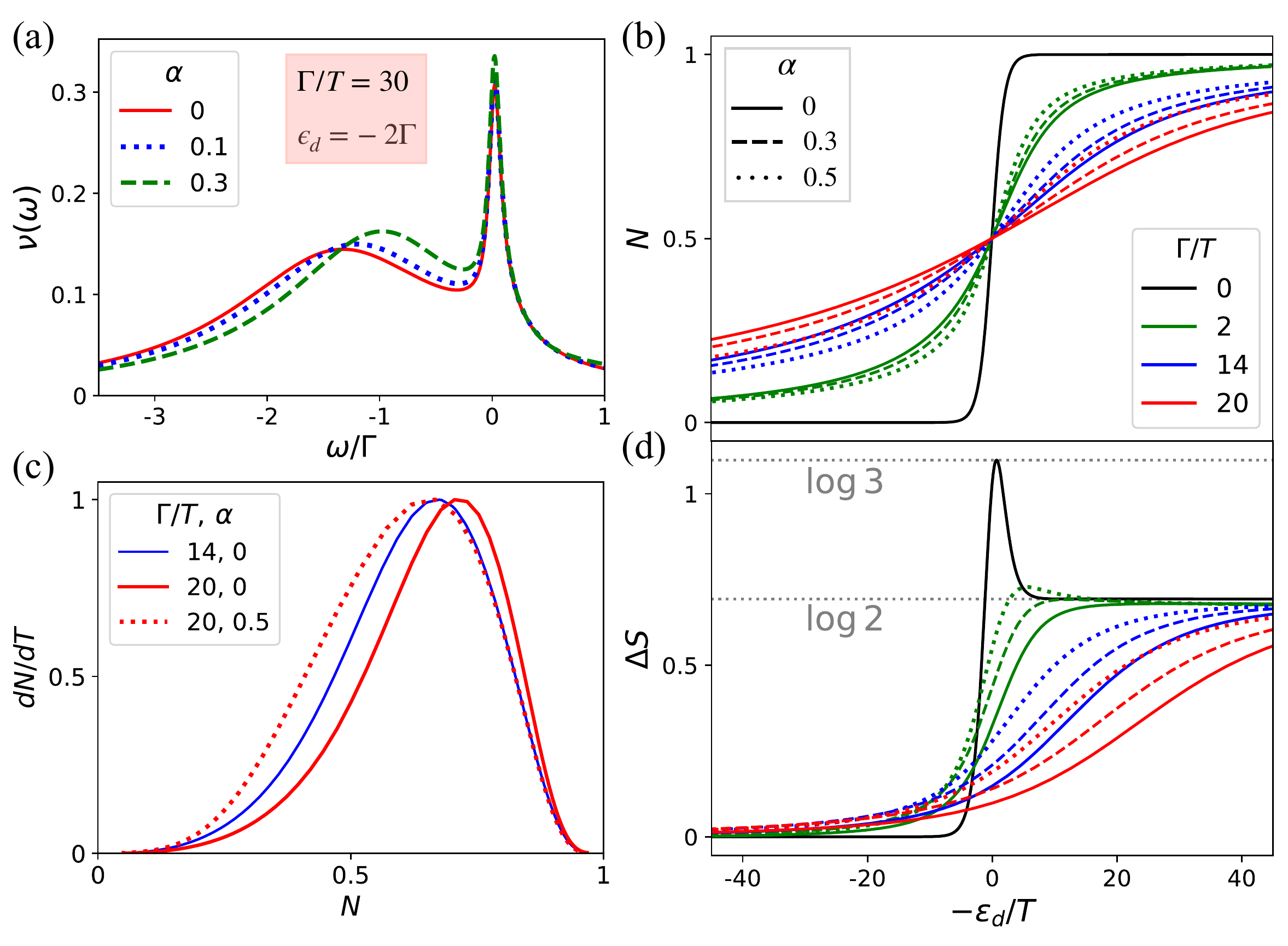}
		\caption{\label{fig:mba_eq} Back action effects due to an unbiased detector. (a)  At large $\Gamma/T$,  the broad peak of the spectral function of the QD level shifts and narrows  with increasing $\alpha$. (b) For any given $\Gamma$, the width of the charge transition narrows with increasing $\alpha$, similar to the effect of reducing $\Gamma$. (c) The peak of $dN/dT$ as a function of $N$ shifts left with increasing $\alpha$, again similar to the effect of reducing $\Gamma$. But the $dN/dT$ peak due to increased $\alpha$ shows an additional broadening compared to that due to a reduced $\Gamma$. The different $dN/dT$ curves here are rescaled to same height for easy comparison. (d) The effect of $\alpha$ on entropy change obtained using the Maxwell relation, for the same set of parameters as in (b). The curves in (b) and (d) are shifted such that $\epsilon_d=0$ corresponds to $N=1/2$.}
	\end{figure}
	
	Figure 2 highlights MBA effects at equilibrium. Numerical renormalization group methods are applied to calculate how two quantities describing the Anderson impurity depend on $\alpha$: the impurity spectral function, $\nu(\omega)$, and its average occupation, $N=\int_{-\infty}^\infty d\omega f(\omega) \nu(\omega)$,  where $f(\omega)$ is the Fermi function. 
	At weak tunneling, $\Gamma/T \ll 1$, $N$ 
	is not affected by $\alpha$ [Fig.~\ref{fig:mba_eq}(b)], even though the spectral function itself broadens with $\alpha$ (not shown)~\cite{SM}. At strong tunneling, $\Gamma/T \gg 1$, $\nu(\omega)$ has a broad peak centered near $\omega=\epsilon_d$, and a Kondo peak at $\omega=0$ [Fig.~\ref{fig:mba_eq}(a)]. The width of the broad peak is $\Gamma$ when $\alpha=0$; it renormalizes downwards as $\alpha$ grows \cite{ma2023identifying, ma2024realizinginteractingresonantlevel}, signaling an effective reduction of the tunnel strength, which consequently reduces the width of the charge transition, $N(\epsilon_d)$, see Fig.~\ref{fig:mba_eq}(b).
	
	It may be difficult to identify the effect of $\alpha$ on $N$ in an experiment, since $\Gamma$ is typically extracted from the charge transition itself. For example, the $\{\Gamma/T=14,\alpha=0\}$ curve in Fig.~\ref{fig:mba_eq}(b) (solid blue) is nearly indistinguishable from the $\{20,0.5\}$ curve (dotted red), which has higher bare $\Gamma/T$ but effective tunnel strength reduced by MBA.  As seen in Fig.~\ref{fig:mba_eq}(c), however, these charge transitions can be more easily distinguished by looking at the temperature-induced modifications to the charge transition,  $dN/dT$.  Although they peak at nearly the same value of $N$, the line shape of $dN/dT(N)$ for $\{20,0.5\}$ is significantly wider than for $\{14,0\}$, by an amount that should be straightforward to distinguish in an experiment. A corollary of the different line shapes of $dN/dT(N)$ is that the entropy change $\Delta S$ obtained by integration of the Maxwell relation, 
	\begin{equation}
		\label{eq:MR}
		\Delta S(\epsilon_d) = \int_{\epsilon_d}^{\infty}\frac{\partial N}{\partial T}|_{\epsilon_d'}d\epsilon_d',
	\end{equation}
	is slightly different for $\{20,0.5\}$  than for  $\{14,0\}$.  We note that the determination of entropy change via the MR is still valid for large $\alpha$ because the system is in thermal equilibrium. In fact, MR measures the entropy change of the combined system plus detector. But for the regime considered here, $\Gamma_q\gg T$, the entropy change of the detector itself is negligible.

	\paragraph{Non-equilibrium MBA effects.}
	We now discuss MBA effects due to a voltage-biased detector. When the QD is 
	decoupled from 
	its lead, 
	$\Gamma \to 0$, the dynamics of the many-body state of the detector is governed by the Hamiltonian $H_n = H_{\rm{det}}+H_{\rm{int}}$. Here $n=\sum_\sigma d^\dagger_\sigma d_\sigma$, the total charge of the system QD, is conserved and is either $0$ or $1$. All the MBA information is  encapsulated in the functions
	\begin{equation}
		\label{eq:a_corr_time}
		A^{+/-}(t)={\rm{Tr}}\left[\rho_{0/1}e^{itH_0}e^{-itH_1}\right],
	\end{equation}
	where  $\rho_n \propto e^{- H_n/T}$  
	denotes the non-equilibrium detector density matrix in which $eV$ is incorporated as a difference of the chemical potential of electrons incoming from source and drain. The resulting functions, $A^\pm (t)$, describe the effect of a sudden change of the detector Hamiltonian that occurs due to a tunnel event into $[A^+(t)]$ or out of $[A^-(t)]$ the QD. These functions can be computed exactly at finite $V$ because both $H_{n=0}$ and $H_{n=1}$ are non-interacting. At long times they decay in an analytically tractable way, as \cite{nozieres1969singularities,aleiner1997dephasing}
	\begin{equation}
		\label{eq:6}
		A^{\pm}(t) \approx  \left(\frac{\pi T}{i\xi \sinh(\pi Tt)}\right)^{\alpha}e^{-\Gamma_{\varphi}t},
	\end{equation}
	where $\xi = \mathcal{O}(\Gamma_{q})$ serves as a high-energy cutoff. While $\alpha$ describes a power-law decay exponent of these functions, the detector bias induces a dephasing rate, $\Gamma_\varphi$. We emphasize that the pair of parameters $\alpha$ and $\Gamma_\varphi$ in Eq.~(\ref{eq:6}) gives a universal characterization of MBA in general detectors. 
	
	\begin{figure}[t]
		\includegraphics[width=\columnwidth]{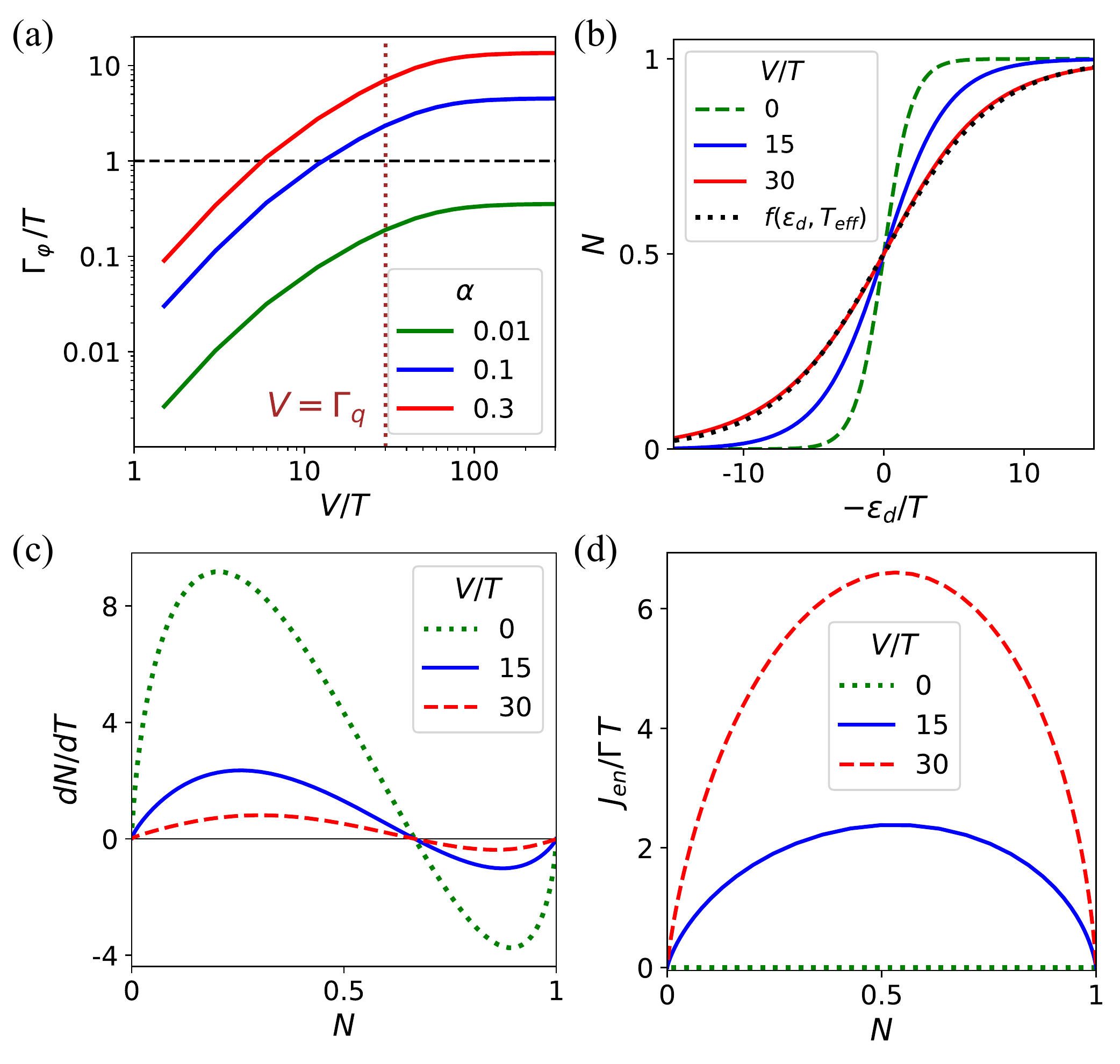}
		\caption{\label{fig:mba_noneq_weaktun} Back action effects due to a biased detector for weak tunneling, $\Gamma \ll T$, at the Anderson impurity. (a) Dependence of the dephasing rate, $\Gamma_\varphi$, on detector bias. Here the line width of the QDD is $\Gamma_q/T=30$. For small $\alpha$, the weak MBA regime of $\Gamma_\varphi \ll T$ is satisified even for $V\gg T$. The case $\alpha=0.3$ is considered in panels (b)-(d). (b) Broadening of the charge transition with detector bias.  The curves are shifted such that $\epsilon_d=0$ corresponds to $N=1/2$. The black dotted line shows that the bias broadened curve for even a  large dephasing rate ($\Gamma_{\varphi}/T \approx 7$) fits nicely to a Fermi function, $f(\epsilon_d, T_{eff})$, with a larger effective temperature ($T_{eff}/T\approx 4$). (c)  Suppression of $dN/dT$ due to increasing detector bias. (d) The dependence of the energy current  from the detector to the system QD, on detector bias and average charge of the QD. 
		} 
	\end{figure}

	Figure~\ref{fig:mba_noneq_weaktun}(a) illustrates the dependence of $\Gamma_\varphi$ on $V$ and the other parameters of the QD detector in our model, using the numerically exact methods for computing $A^\pm(t)$ and its Fourier transforms $A^\pm(\omega)$ reported in our previous work \cite{sankar2024detector}. We find that $\Gamma_\varphi \propto V$ for $T  \lesssim  V \lesssim \Gamma_{q}$ and it saturates for $V > \Gamma_{q}$. $\Gamma_\varphi$ itself increases with $\alpha$ and vanishes when $\alpha \to 0$. Importantly, for small $\alpha$, the weak MBA regime of $\Gamma_\varphi \ll T$ is satisfied even for $V\gg T$. 
	
	We now discuss the distinct implications of the dephasing rate on $N$ and $dN/dT$ in the weak tunneling case by treating finite $\Gamma$ within a rate equations approach~\cite{sankar2024detector,SM}. The width of the charge transition grows with $V$ as soon as $\Gamma_\varphi$ is larger than $T$ [Fig.~\ref{fig:mba_noneq_weaktun}(b)].  In this way, the presence of strong dephasing, $\Gamma_\varphi \gtrsim T$, can be easily detected in experiments, in the form of bias-dependent broadening of the charge transition. Because charge transitions broadened primarily by $\Gamma_\varphi$ are consequently less affected by $T$, $dN/dT$ is suppressed at high detector biases [Fig.~\ref{fig:mba_noneq_weaktun}(c)].
	This signals that the system is out of thermal equilibrium, and the application of thermodynamic relations is made unreliable by MBA. Interestingly, even when $\Gamma_\varphi \gtrsim T$,  the broadened charge transition  curves fit very well to thermal broadening lineshapes, but with a larger effective temperature $T_{\rm{eff}}>T$, as shown in Fig.~\ref{fig:mba_noneq_weaktun}(b). As a result, a naive application of the MR can yield an entropy change that is approximately correct, albeit with a rescaled temperature $T_{\rm{eff}}(T)$ of the QD lead. 
	
	\paragraph{Detector non-crossing approximation (D-NCA).}  
	To treat non-equilibrium MBA effects at a stronger tunnel coupling, we develop an
	approximate non-perturbative
	method in which a hybridization expansion is performed in the slave-boson 
	framework resorting to the non-crossing approximation (NCA), with the MBA information entering as modified bare (i.e. $\Gamma\to 0$) QD propagators, see Fig.~\ref{fig:setup}(b). 
	We now briefly describe this method. A more detailed description is provided in \cite{SM}. 
	
	In the slave-boson formalism, a different Hilbert space is considered for the system QD, including a (non-physical) vacuum $ |\Omega\rangle $, an empty state created by the slave-boson operator $ b^\dagger  |\Omega\rangle =|0\rangle $, and the singly occupied states created by the pseudo-fermion operators $f_\sigma^\dagger | \Omega\rangle =|\sigma\rangle$. The original fermion creation operators are expressed in terms of these new operators as  $d_\sigma^\dagger =f_\sigma^\dagger b$. The expression of $H_{\rm{sys}}$ becomes
	\begin{equation}
		H_{\rm{sys}}^{\rm SB}=\sum_\sigma\epsilon_d f^\dagger_\sigma f_\sigma +\sum_{k\sigma} \epsilon_k \hat{n}_{k\sigma} +  \sum_{k \sigma} \gamma_k \left(c^\dagger_{k\sigma}b^\dagger f_\sigma +{\rm H.c.} \right).
	\end{equation}
	For $\lambda=0$, the NCA is known to be highly accurate, even well below the Kondo temperature \cite{bickers1987review,cox1987properties}. 
	Under NCA, tunneling is treated using a  simple Dyson-like diagrammatic expansion in $\Gamma$~\cite{wingreen1994anderson}. Our detector-NCA (D-NCA) approach generalizes the  NCA treatment to $\lambda\ne 0$. A key difference is that the bare retarded and lesser pseudo-fermion propagators  are modified due to the interaction with the detector using the $A^\pm$ functions,
	\begin{equation}
		G_{f\sigma,0}^r(\omega)=-iA^{+,r}(\omega-\epsilon_d),\, G_{f\sigma,0}^<(\omega)=i A^-(\omega-\epsilon_d),
	\end{equation}
	where $A^{+,r}(\omega)$ is the Fourier transform of $\theta(t)A^+(t)$, with $\theta(t)$ denoting the Heaviside step function.  
	\begin{figure}[t]
		\includegraphics[width=\columnwidth]{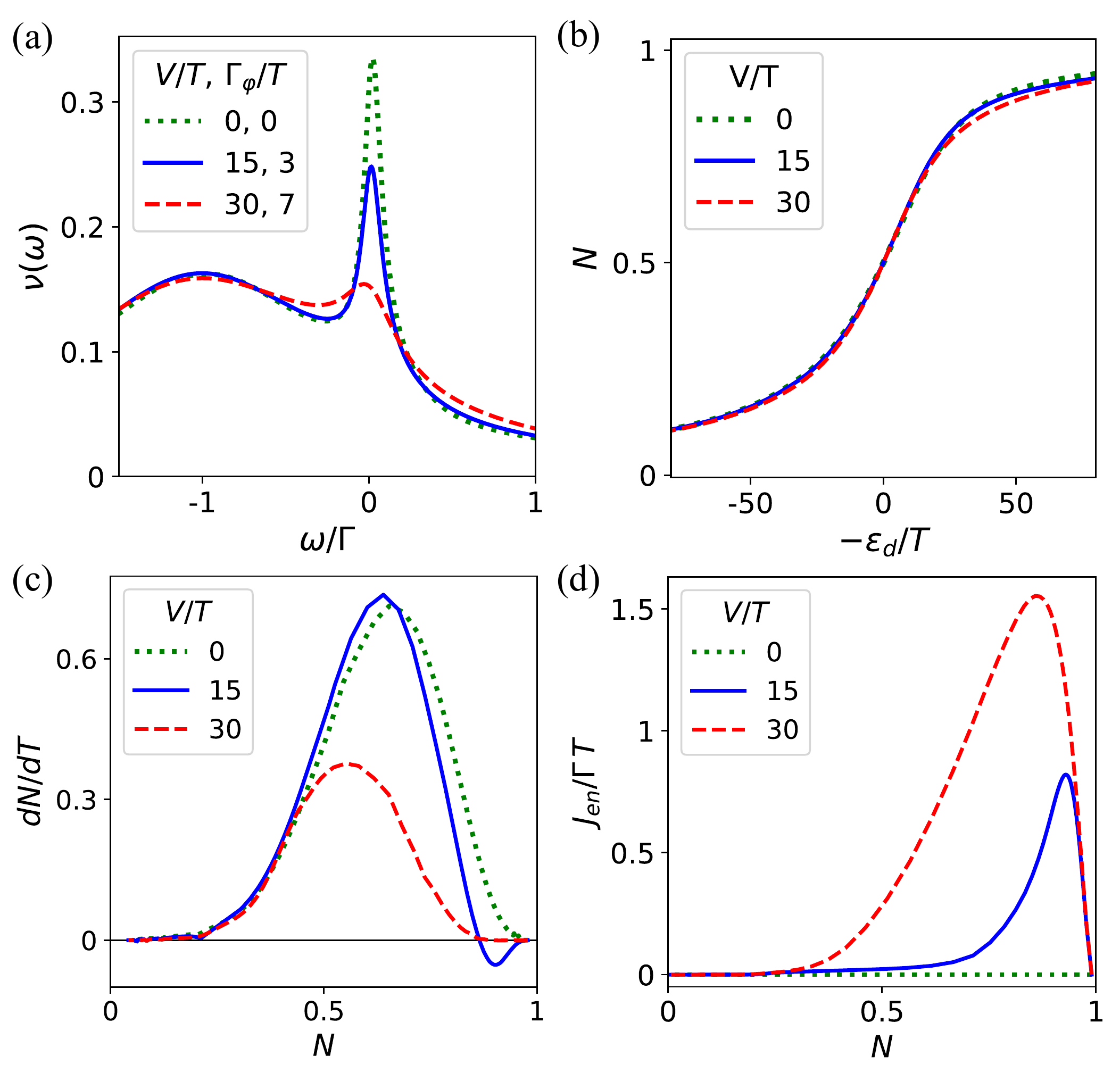}
		\caption{\label{fig:mba_noneq_strongtun}
			Back action effects ($\alpha=0.3$) due to a biased detector for the case of strong tunneling,  $\Gamma/T=30$. (a) The suppression of the Kondo peak at $\epsilon_d=-2\Gamma$ with detector bias. (b) The bias has a very weak effect on the impurity charge for values of $V/T$ shown, each of which satisfy $\Gamma_\varphi \ll \Gamma$. The curves here are shifted such that $\epsilon_d=0$ corresponds to $N=1/2$. (c) $dN/dT$ gets suppressed with increasing bias, but only for $N\gtrsim 1/2$. (d) The energy current that flows from the detector to the system is skewed with the peak occurring at $N$ close to $1$.
		} 
	\end{figure}
	
	The additional approximation in D-NCA is neglecting higher order mixed irreducible self-energies with both interaction and tunneling vertices, similar in spirit to NCA. The D-NCA approximation scheme becomes close to exact in two regimes: (i) $\Gamma \ll T$ together with arbitrary $\alpha$, and (ii) $\alpha \ll 1$ together with arbitrary $\Gamma$. The D-NCA is expected to be a decent approximation for the intermediate regimes that are discussed in the paper, but might fail in the extreme simultaneous regime of both $\alpha \gg 1$ and $\Gamma \gg T$. 
	
	\paragraph{MBA in the Kondo regime.}  
	We now proceed to discuss the results obtained using D-NCA in the strong tunneling regime, see Fig.~\ref{fig:mba_noneq_strongtun}. 
	As in Fig.~\ref{fig:mba_eq}(a), a Kondo peak appears in $\nu(\omega)$ for $\Gamma \gg T$.  The peak is suppressed with increasing detector bias, similar to the findings of Ref.~\cite{silva2003peculiarities}, and as observed experimentally in Ref.~\cite{avinun2004controlled}. The width of the charge transition is only weakly affected by the bias voltage because it is dominated by tunneling, $\Gamma \gg \Gamma_\varphi$ [Fig.~\ref{fig:mba_noneq_strongtun}(b)]. Of particular interest is the behavior of $dN/dT$ in Fig.~\ref{fig:mba_noneq_strongtun}(c). At $V=0$ the peak in $dN/dT$ occurs at $N>0.5$, consistent with the rightward shift of the peak in Fig.~\ref{fig:mba_eq}(c). Upon increasing bias, $dN/dT$ is suppressed, but in an $N-$dependent manner so the lineshape is changed.  This contrasts with the weak tunneling case in Fig.~\ref{fig:mba_noneq_weaktun}(c), where $dN/dT$ versus $N$ is rescaled by a constant factor but does not change shape.
	
	The essential conclusion from this analysis, including both the weak and strong tunneling cases, is that the necessary and sufficient measurement condition to minimize MBA for thermodynamic measurements is to ensure that $\Gamma_\varphi \ll T$.  In the case of weak coupling, it is possible to make meaningful measurements even when $\Gamma_\varphi \gtrsim T$ through the use of an effective temperature, but in the case of strong coupling, even that is not possible.
	
	\paragraph{Energy flow.} 
	
	To gain further insight into the difference in the nature of the suppression of $dN/dT$ between weak and strong tunneling regimes, we calculate the energy flow from the non-equilibrium detector to the impurity QD. As a proxy to this quantity, we note that away from equilibrium the impurity distribution function differs from the Fermi function, and the difference between these is related to heat flow.  At equilibrium, the lesser Green's function satisfies $G^<(\omega) = 
	2 \pi i \nu(\omega) 
	f(\omega)$.  We define the non-equilibrium distribution function of the QD---which determines how the available states are occupied---as 
	$h(\omega)\equiv i G^<(\omega)/(2\pi\nu(\omega))$. The energy current from the QD to the lead  can be expressed as \cite{SM}
	\begin{equation}
		\label{eq:en_cur}
		J_{\rm{en}}=2\Gamma  \int_{-\infty}^{\infty}d\omega\,\omega\, \nu(\omega)(h(\omega)-f(\omega)).
	\end{equation}
	The factor of 2 accounts for spin. In the steady state, the energy flow from the detector to the system's QD equals the energy flow from the system's QD to its lead, so Eq.~(\ref{eq:en_cur}) also represents the heat flow between detector and QD.

	We present results for heat flow based on D-NCA in Fig.~\ref{fig:mba_noneq_weaktun}(d) and Fig.~\ref{fig:mba_noneq_strongtun}(d), representing weak and strong tunneling respectively.  The quantity $\frac{J_{\rm{en}}}{\Gamma T}$ that is plotted represents how far the system is from thermal equilibrium.
	As long as $\Gamma_{\varphi} \ll T$, the system stays close to thermal equilibrium and $\frac{J_{\rm{en}}}{\Gamma T} \ll 1$ for all N. When $\Gamma_\varphi \gtrsim T$, the energy flow as a function of the average QD charge is qualitatively different for weak and strong tunnel couplings [Fig.~\ref{fig:mba_noneq_weaktun}(d) compared to Fig.~\ref{fig:mba_noneq_strongtun}(d)].  For weak tunneling, it peaks at $N\sim 1/2$ as might naively be expected given that charge fluctuations in a weakly-coupled QD would also be peaked at $N=1/2$.  In contrast, for strong tunneling, heat flow is large only  close to $N=1$.
	
	While we currently do not have a simple picture for the anomalous $N$ dependence of the energy flow, we note that this dependence is similar to the corresponding $N$-dependent suppression of $dN/dT$, which is also peaked at $N$ close to 1.  At an intuitive level, this may be understood from the fact that both quantities directly relate to the extent of  deviation of the system from thermal equilibrium.   An experiment probing $N$-dependent heat flow directly in the Anderson impurity-detector setup would be especially insightful in developing a better understanding of the mechanism.

	\paragraph{Acknowledgements.} We gratefully acknowledge support from the European Research Council (ERC) under the European Union Horizon 2020 research and innovation programme under grant agreement No. 951541. ARO (W911NF-20-1-0013). Y.M. acknowledges support from ISF Grants Nos. 359/20 and 737/24. J.F. acknowledges support from the Natural Sciences and Engineering Research Council of Canada; the Canadian Institute for Advanced Research; the Max Planck-UBC-UTokyo Centre for Quantum Materials and the Canada First Research Excellence Fund, Quantum Materials and Future Technologies Program.  We thank  Zhanyu Ma and Cheolhee Han for illuminating discussions. 
	Numerical renormalization group calculations where carried out with the \QSpace tensor library \cite{weichselbaumNonabelianSymmetriesTensor2012,weichselbaumXsymbolsNonAbelianSymmetries2020,weichselbaumQSpaceOpensourceTensor2024}.

\end{document}


\title{Supplement: Back-action effects in charge detection}
	
	\author{Sarath Sankar}
	\affiliation{School of Physics and Astronomy, Tel Aviv University, Tel Aviv 6997801, Israel}
	\author{Matan Lotem}
	\affiliation{Department of Physics, Ben-Gurion University of the Negev, Beer-Sheva 84105, Israel}
	\affiliation{Department of Physics, Princeton University, Princeton, NJ 08544, USA}
	\author{Joshua Folk}
	\affiliation{Quantum Matter Institute, University of British Columbia, Vancouver, British Columbia, Canada}
	\affiliation{Department of Physics and Astronomy, University of British Columbia, Vancouver, British
		Columbia, Canada}
	\author{Eran Sela}
	\affiliation{School of Physics and Astronomy, Tel Aviv University, Tel Aviv 6997801, Israel}
	\author{Yigal Meir}
	\affiliation{Department of Physics, Ben-Gurion University of the Negev, Beer-Sheva 84105, Israel}
	
	\maketitle
	In this supplementary material we provide a detailed description of our method, D-NCA, which incorporates the MBA functions $A^\pm(\omega)$ into the standard Green-function based NCA~\cite{wingreen1994anderson}. We consider throughout the model in Eqs.~(1-3) of the main text.
	
	\section{\label{sec:exact_weak_tun}Exact treatment of MBA at weak tunnel strength}
	
	At weak tunnel strength, the  functions $A^\pm(\omega)$ modify the tunnel rates in and out of the QD. 
	They are given by~\cite{sankar2024detector}
	\begin{eqnarray}
		\Gamma_{\rm{in}} (\epsilon_d)&=&\int_{-\infty}^{\infty} d\omega\, A^+(\omega-\epsilon_{d})f(\omega)\label{eq:tun_in_rate},\\
		\Gamma_{\rm{out}}(\epsilon_d) &=& \int_{-\infty}^{\infty} d\omega\, A^-(\omega-\epsilon_{d})f(-\omega). \label{eq:tun_out_rate}
	\end{eqnarray}
	Physically, the functions $A^{\pm}(\omega)$ gives the probability of the detector
	to absorb/emit energy $\omega$ in the tunneling-in/out process of the QD.
	
	Let us determine the occupation of the QD in the weak tunneling regime, $\Gamma \ll T$. Let $P_{0}, \, P_{\sigma}$ denote the occupation probability of the empty, and singly occupied states with spin $\sigma$, with $\sum_\sigma P_\sigma + P_0=1$. These probabilities are determined by the rate equation in steady state,
	\begin{equation}
		\label{eq:steady-state}
		\frac{dP_\sigma}{dt} = P_0 \Gamma_{\rm{in}} - P_\sigma \Gamma_{\rm{out}} =0.
	\end{equation}
	The average occupation, $N=\sum_\sigma P_\sigma$, is then given by
	\begin{equation}
		N(\epsilon_d)=\frac{2\Gamma_{\rm{in}}(\epsilon_d)}{2\Gamma_{\rm{in}}(\epsilon_d)+\Gamma_{\rm{out}}(\epsilon_d)}.
	\end{equation}
	Let us introduce the exact retarded, lesser and greater Green functions of the $d-$ level in presence of both tunneling to the lead and interaction with the detector, defined as,
	\bea
	G^r_\sigma &=& -i \theta(t) \langle \{ d_\sigma(t),d_\sigma^\dagger(0) \} \rangle, \nonumber \\
	G^>_\sigma &=& -i \langle  d_\sigma(t)d_\sigma^\dagger(0)  \rangle, \nonumber \\
	G^<_\sigma &=& i \langle  d_\sigma^\dagger(0) d_\sigma(t) \rangle,
	\eea
	along with the corresponding Fourier transform $G^{r,<,>}(\omega)=\int_{-\infty}^\infty e^{i \omega t} G^{r,<,>}(t)$. We also define the spectral function of the $d-$level 
	\be
	\label{eq:dos_def}
	\nu_{\sigma}(\omega)=-\frac{1}{\pi} {\rm{Im}}\left(G^r_\sigma(\omega)\right).
	\ee
	
	We define by $G^{r,>,<}_{\sigma,0}$ the bare Green functions with respect to $\Gamma$, which however take into account exactly the interaction with the detector $\lambda$. We find that
	\be
	\label{eq:bare_less_great}
	G^{>/<}_{\sigma,0}(\omega;\epsilon_d)=\mp iP_{0/\sigma}(\epsilon_d)A^\pm(\omega-\epsilon_d), 
	\ee
	and the spectral function of the $d-$level can be expressed as~\cite{sankar2024detector} $\nu_{\sigma,0}(\omega;\epsilon_d)=\frac{1}{2\pi} (P_0(\epsilon_d) A^+(\omega-\epsilon_d) + P_\sigma(\epsilon_d)A^-(\omega-\epsilon_d))$.
	
	These relations can be proven using a Lehmann representation as in Ref.~\cite{sankar2024detector}. Let us show this explicitly for the greater Green function. Let $|\psi^n_i\rangle$ denote the eigenbasis of $H_n$, where $H_n$ is the detector Hamiltonian which incorporates $\lambda$, depending on the occupation of the QD $n=0,1$. The eigenenergies of these many body states are denoted $E_i^n$. The density matrix of the detector starting at a particular $n$ is described by $P_i^n=e^{-\beta E_i^n}/(\sum_j e^{-\beta E_j^n})$. Then,
	\bea
	G^{>}_{\sigma,0}(\omega;\epsilon_d)&=&-i \int dt e^{i \omega t}   \langle d_\sigma(t)d^\dagger_\sigma(0) \rangle \nonumber \\
	&=&-i \int dt e^{i \omega t} \sum_{i,j}P_0  P_j^0 | \langle \psi_j^0| \psi_i^1 \rangle |^2 e^{-i \epsilon_d t} e^{i E_j^0 t} e^{-i E_i^1 t} = -i P_0 A^+(\omega-\epsilon_d),
	\eea
	where we used the Lehmann representation of $A^+(\omega)$ given in Eq.~(A7) in Ref.~\cite{sankar2024detector}. 
	
	\subsection{$V=0$}
	
	At thermal equilibrium, the $A^\pm(\omega)$ correlators are related by the fluctuation-dissipation theorem (FDT),
	\begin{equation}
		A^-(\omega)=e^{-\beta (\omega+\Delta(T))} A^+(\omega),
	\end{equation}
	where $\Delta(T)=F_{\rm{det}}(n=0)-F_{\rm{det}}(n=1)$ and $F_{\rm{det}}(n)$ denotes the Free energy of the detector that depends on the charge state, $n$, of the system QD. The free energy difference $\Delta$ in general is temperature dependent. 
	
	This again follows from the Lehmann representation,
	\bea
	A^-(\omega)&=&\int dt {\rm{Tr}} \rho_1 e^{i t H_0}e^{-i t H_1} e^{i \omega t} \nonumber \\ 
	&=&\int dt \sum_{i,j} \frac{e^{-\beta E_j^1}}{Z_1}| \langle \psi_j^0| \psi_i^1 \rangle |^2 e^{i t E_j^0}e^{-i t E_i^1} e^{i \omega t} \nonumber \\ 
	&=&\sum_{i,j} \frac{e^{-\beta E_j^1}}{Z_1} | \langle \psi_j^0| \psi_i^1 \rangle |^2 2 \pi \delta(E_j^0-E_i^1 + \omega)
	\nonumber \\ 
	&=&e^{-\beta \omega} \sum_{i,j} \frac{e^{-\beta E_i^0}}{Z_1} | \langle \psi_j^0| \psi_i^1 \rangle |^2 2 \pi \delta(E_j^0-E_i^1 + \omega)\nonumber \\ 
	&=&e^{-\beta \omega} \frac{Z_0}{Z_1}\sum_{i,j}  A^+(\omega).
	\eea
	
	Then from Eqs.~\eqref{eq:tun_in_rate}, \eqref{eq:tun_out_rate}, it follows that $\Gamma_{\rm{out}}(\epsilon_d)=e^{\beta (\epsilon_d -\Delta(T) )}\Gamma_{\rm{in}}(\epsilon_d)$ and consequently,
	\begin{equation}
		\label{eq:NDelta}
		N(\epsilon_d)=\frac{1}{1+\frac{1}{2}e^{\beta (\epsilon_d-\Delta(T))}}.
	\end{equation}
	We can also arrive at the above result from 
	the total partition function of the system and detector,
	\begin{equation}
		Z=e^{-\beta F_{\rm{det}}(n=0)}+2e^{-\beta \epsilon_d}e^{-\beta F_{\rm{det}}(n=1)}.
	\end{equation}
	Eq.~(\ref{eq:NDelta}) immediately follows.
	
	Thus the average occupation in the weak tunneling limit is not affected by MBA at thermal equilibrium other than an overall shift, which can be referred to as a Hartree shift. The temperature dependence of $\Delta(T)$ would bring in the entropy change of the detector between $n=0$ and $n=1$ when the entropy change is calculated using the Maxwell relation. In the main text, we consider the level width of the QDD to be much greater than temperature, $\Gamma_q\gg T$, and in this case the entropy change of the detector is negligible.
	
	\subsection{$V \ne 0$}
	At equilibrium $G^<(\omega)=-2i f(\omega) {\rm{Im}}G^r(\omega)$. This motivates the definition 
	\be
	\label{eq:h_def}
	h_{\sigma}(\omega)  \equiv    - i G^<_{\sigma}(\omega)/(2\pi\nu_{\sigma}(\omega)),
	\ee
	for the non-equilibrium distribution function. In the weak tunneling case, using the Eq.~\eqref{eq:bare_less_great}, we get,
	\begin{equation}
		h_\sigma(\omega)=\frac{1}{1+\frac{P_0(\epsilon_d)A^+(\omega-\epsilon_d)}{P_\sigma(\epsilon_d)A^-(\omega-\epsilon_d)}}.
	\end{equation}
	Note that if the FDT is obeyed then $h_\sigma(\omega)=f(\omega)$. 
	From the long-time analytical expression of these correlators in the manuscript, we see that the deviation from the FDT happens when $\Gamma_{\varphi}\gtrsim T$. 
	
	\section{Detector- Non-crossing approximation (D-NCA)}
	The details of the D-NCA method are discussed in this section. The main idea is to perform an hybrdization expansion for the propagators of the $d$-level, like standard NCA~\cite{wingreen1994anderson}, in the slave boson framework, by starting with bare propagators ($\Gamma = 0$) that are modified to include MBA from the detector in an exact manner.

	\subsection{Slave boson representation and MBA-modified bare propagators}
	For the sake of completeness, we  review the key definition of the problem in terms of Green functions using  the slave-boson (SB) representation, closely following Ref.~\cite{wingreen1994anderson}.The discussion of the MBA modified bare propagators occurs towards  the end of this subsection, in Eq.~(\ref{eq:bare_fermion}).
	
	Since we take $U$ to be infinite, the Hilbert space of the $d$-level   consists of 3 states: $|0\rangle , \, |\uparrow\rangle ,\, |\downarrow\rangle$. In the SB formalism, a different Hilbert space is considered, generated from a nonphysical vacuum $ |\Omega\rangle $, using a bosonic creation operator $b^\dagger \rightarrow b^\dagger  |\Omega\rangle =|0\rangle $, and  fermionic creation operators, $f_\sigma^\dagger \rightarrow f_\sigma^\dagger | \Omega\rangle =|\sigma\rangle$. Additional states such as $f^\dagger_\sigma b^\dagger  |\Omega\rangle$ are nonphysical. The original fermionic operators are expressed in terms of these new operators as
	\begin{equation}
		d_\sigma =b^\dagger f_\sigma,\quad d_\sigma^\dagger =f_\sigma^\dagger b.
	\end{equation}
	
	One can write the full Hamiltonian of the model given in Eqs.~(1-3) of the main text in terms of these new operators,
	
	\begin{equation}
		H^{\rm SB} = H_{\rm{sys}}^{\rm SB}+ H_{\rm{det}}+ H_{\rm{int}}^{\rm SB},
	\end{equation}
	where the system and interaction part of the Hamiltonian become
	\begin{eqnarray}
		H_{\rm{sys}}^{\rm SB}&=& \epsilon_d \sum_\sigma f^\dagger_\sigma f_\sigma + \sum_{k\sigma} \epsilon_k \hat{n}_{k\sigma} + \sum_{k \sigma} \left( \gamma_k c^\dagger_{k\sigma}b^\dagger f_\sigma + {\rm H.c.} \right), \\
		H_{\rm{int}}^{\rm SB}&=& \lambda \sum_{i=1}^m\left(q^\dagger_i q_i -\frac{1}{2}\right)\left(\sum_\sigma f^\dagger_\sigma f_\sigma -\frac{1}{2}\right).
	\end{eqnarray}
	
	The advantage of the SB formalism is that $H^{\rm SB}$ does not allow double occupation of the QD, since $d_\sigma^\dagger |\bar{\sigma}\rangle = f_\sigma^\dagger b |\bar{\sigma}\rangle =0$. Thus it automatically enforces the $U\to \infty$ condition. 
	The new complication that arises in the SB formalism is that now there is a constraint: the total number of particles, $Q=b^\dagger b + \sum_\sigma f_\sigma^\dagger f_\sigma $, should be restricted to unity.  It is this subspace which coincides with  the original Hilbert space of the $d$-level. Notice that even though $H^{\rm SB}$ conserves $Q$, one still should project to the particular $Q=1$ subspace. One solution is to first work in a unconstrained grand-canonical ensemble of the $b$ and $f_\sigma$ particles with a chemical potential, here denoted by $i\mu$, and finally impose the constraint $Q=1$ by a Fourier transformation with respect to the chemical potential.  We define averages in this grand-canonical ensemble as \begin{equation}
		\langle O \rangle_{i\mu} =\frac{1}{Z_{i\mu}} {\rm{Tr}} \left[e^{-\beta (H^{\rm SB}+i\mu Q)}T_C\left[S_C(-\infty, -\infty )O\right] \right],
	\end{equation}
	where the trace  includes summation over different $Q$ sectors. Here $T_C$ in a time ordering operator along the Keldysh contour, $C$, and 
	\begin{equation}
		S_C (\infty,-\infty)=e^{-i\int_C dt'\, H^{\rm SB}(t')}.
	\end{equation} These grand-canonical averages can be computed using standard Green function methods.
	
	We are interested in particular operators of the original fermions whose expectation value vanish in the $Q=0$ ensemble, for example $O=d^\dagger_\sigma d_\sigma(t)$. For such operators, one can relate the desired averages in the $Q=1$ subspace to grand-canonical averages, as follows, see Eq.~(\ref{eq:oper_relation}). Since $Q$ is conserved, the grand canonical average and the partition function can be written as
	\begin{eqnarray}
		\langle O \rangle_{i\mu} &=&\frac{1}{Z_{i\mu}}\left[ \sum_{q=0}^{\infty}e^{-i\beta \mu q}Z_{Q=q} \langle O \rangle_{Q=q}  \right] 
		\label{eq:O_q_expansion},\\
		Z_{i\mu}&=& \sum_{q=0}^{\infty} e^{-i\beta \mu q}Z_{Q=q}. \label{eq:Z_q_expansion}
	\end{eqnarray}
	Thus, one can write grand-canonical averages as
	\be
	\langle O \rangle_{i\mu}=\frac{\sum_{q=0}^{\infty}e^{-i\beta \mu q}Z_{Q=q} \langle O \rangle_{Q=q} }{Z_0+Z_1 e^{-i \beta \mu}+\dots }.
	\ee
	Consider expanding this expression in powers of $ e^{-i \beta \mu}$ and denote the coefficient of $ e^{-i q\beta \mu}$ as $\langle O \rangle_{i\mu}^{(q)}$. It is evident that the coefficient of $e^{-i \beta \mu}$ is
	\be
	\langle O \rangle_{i\mu}^{(1)}=\frac{Z_1}{Z_0} \langle O \rangle_{Q=1},
	\ee
	where we used $\langle O \rangle_{Q=0}=0$. Hence,
	\begin{eqnarray}
		\langle O \rangle_{Q=1}=\frac{Z_{Q=0}}{Z_{Q=1}} \langle O \rangle_{i\mu}^{(1)}, \label{eq:oper_relation}
	\end{eqnarray}
	
	
	The expression for the ratio of partition functions in the RHS of Eq.~\eqref{eq:oper_relation} can be obtained by putting $O=Q=1$, yielding 
	\begin{equation}
		\frac{Z_{Q=1}}{Z_{Q=0}}= \langle b^\dagger b\rangle_{i\mu}^{(1)} +\sum_\sigma\langle f_\sigma^\dagger f_\sigma \rangle_{i\mu}^{(1)} .
	\end{equation}
	
	Let us define the following projected Green functions for the pseudo fermions and slave boson,
	\begin{eqnarray}
		D^r(t)&=&-i\, \theta(t)\langle [b(t),b^\dagger(0)]\rangle_{i\mu}^{(0)},\quad 	D^>(t)=-i \langle b(t)b^\dagger(0)\rangle_{i\mu}^{(0)},\quad D^<(t)=-i \langle b^\dagger(0)b(t)\rangle_{i\mu}^{(1)}, \label{eq:boson_gfn} \\
		G_{f\sigma}^r(t)&=&-i \, \theta(t)\langle \{f_\sigma(t),f_\sigma^\dagger(0)\}\rangle_{i\mu}^{(0)},\quad G_{f\sigma}^>(t)=-i \langle f_\sigma(t)f_\sigma^\dagger(0)\rangle_{i\mu}^{(0)},\quad	G_{f\sigma}^<(t)=i\langle f_{\sigma}^\dagger(0)f_\sigma(t)\rangle_{i\mu}^{(1)}, \label{eq:fermion_gfn}
	\end{eqnarray}
	where $[\,]$ denotes commutator and $\{\,\}$ denotes anti-commutator. Note that the different Green functions are defined as differnet coefficient with respect to the expansion in powers of $e^{-i \beta \mu}$. Particularly the retarded and greater functions are defined as the $O(1)$ coefficient ($Q=0$ ensemble) while the lesser functions are defined as the $O(e^{-i\beta \mu})$ coefficients ($Q=1$ ensemble). The reason behind this particular choice  will be clarified below. Clearly,  if one considers the lesser function  in the $Q=0$ subspace, it will trivially vanish because of the annihilation operator on the right. 
	For the same reason, we also have
	\begin{equation}
		\label{eq:retarded_greater_relation}
		D^r(t)=\theta(t) D^>(t),\quad  G_{f\sigma}^r(t)=\theta(t) G_{f\sigma}^>(t) \quad \implies D^>(\omega) = 2i \,{\rm{Im}}\left[D^r(\omega)\right], \quad G_{f\sigma}^>(\omega)=2i\,{\rm{Im}}\left[G_{f\sigma}^r(\omega)\right],
	\end{equation}
	with the last two equations referring to the respective Fourier transforms.
	
	So far the Green functions are defined as the exact Green functions. We now define the Green functions in the limit 
	$\Gamma = 0$ as the ``bare" Green functions, and denote them with a  subscript $0$. Since the boson operators only enter $H^{\rm SB}$ in the  tunneling terms, the bare projected 
	boson Green functions are trivial,
	\begin{eqnarray}
		\label{eq:bare_pseudo}
		D_0^r(\omega)=\frac{1}{\omega+i\eta} \quad ,~~~~~~~ \quad D_0^<(\omega)=-2\pi i P_0\delta (\omega),
	\end{eqnarray}
	where $P_0$ denotes the probability of the empty state. The bare projected pseudo-fermion  Green  functions can  be easily written in terms of the $A^\pm$ correlators 
	by substituting $H_n$ with $H_n^{\rm SB}=H_{\rm{int}}^{\rm SB}+H_{\rm{det}}$. We then get
	\begin{eqnarray}
		\label{eq:bare_fermion}
		G_{f\sigma,0}^r(\omega)=-i A^{+,r}(\omega-\epsilon_d), \quad 
		G_{f\sigma,0}^<(\omega)=i P_{\sigma}A^-(\omega-\epsilon_d),
	\end{eqnarray}
	where $ P_\sigma$  denotes the  probability of the singly occupied state with spin $\sigma$. 
	The bare form of the projected greater Green function follows from Eq.~\eqref{eq:retarded_greater_relation}. Note that since the projected retarded and greater Green functions are defined in the $Q=0$ ensemble, there are no occupation probabilities associated with it. We will see later that after the hybridization expansion, the information of the initial occupation probabilities in the bare form of the projected lesser Green functions will be erased. 
	
	\subsection{NCA and hybridization expansion}
	A primary quantity that we wish to calculate is the retarded Green function of the $d$-level,
	\begin{equation}
		G^r_{\sigma}(t)=-i\theta(t)\left< \left\{d_{\sigma}(t),d_{\sigma}^\dagger(0)\right\}\right> = -i\theta(t)\left< \left\{b^\dagger(t) f_{\sigma}(t),f_{\sigma}^\dagger(0)b(0)\right\}\right>_{Q=1}= 	\frac{Z_{Q=0}}{Z_{Q=1}}G_{\sigma,i\mu}^{r(1)}(t) \label{eq:g_reta_qd}.
	\end{equation}
	To proceed we  make two approximations: (1) neglect vertex corrections in Eq.~\eqref{eq:g_reta_qd} factoring it out into a product of  bosons and fermion Green functions, and (2) keep only the leading contribution to the irreducible self-energies. These two are the approximations used in the standard NCA treatment in the absence of detector. Approximation (2) in the standard NCA amounts to discarding higher order self-energy diagrams with crossing reservoir lines.
	In D-NCA, the  diagrams discarded due to approximation (2) also include higher order mixed irreducible self-energies with both interaction and tunneling vertices. Applying approximation (1), we obtain 
	\begin{equation}
		\label{eq:dnca_factorization}
		G_{\sigma, i\mu}^{r(1)}(t) \overset{\rm{D-NCA}}{=} -i \theta(t)\left[D^>(-t)G_{f\sigma}^<(t)-D^<(-t)G_{f\sigma}^>(t)\right],
	\end{equation}
	which involves the projected Green functions defined in Eq.~\eqref{eq:boson_gfn} and Eq.~\eqref{eq:fermion_gfn}.
	The reason for the particular choice of the projected Green functions and why only the kind of factorization in Eq.~\eqref{eq:dnca_factorization} survives after applying approximation (1) is simple: by definition, 	$G_{\sigma, i\mu}^{r(1)}(t)$ is of $O(e^{-i\beta \mu})$, which factorizes to include one $O(e^{-i\beta \mu})$  factor and another $O(1)$ factor. The $O(1)$ part of the lesser functions trivially vanishes, and hence the only combination that survives are the ones in Eq.~\eqref{eq:dnca_factorization}.
	
	The next task is to obtain a diagrammatic expansion of the projected boson and pseudo-fermion Green functions in terms of the bare functions. This can  be done by considering the diagrammatic expansion in the unconstrained grand-canonical ensemble and then keeping only terms to the required order by simple power counting.
	For the  diagrammatic expansion in the grand-canonical ensemble we resort to the approximation (2). We will see that the projected Green functions are easily obtained from such a  diagrammatic expansion by simply replacing the usual Green function lines with the projected ones. This will be more apparent when we discuss the Dyson equations.
	
\begin{figure*}
		\centering
		\includegraphics[width=0.9\textwidth]{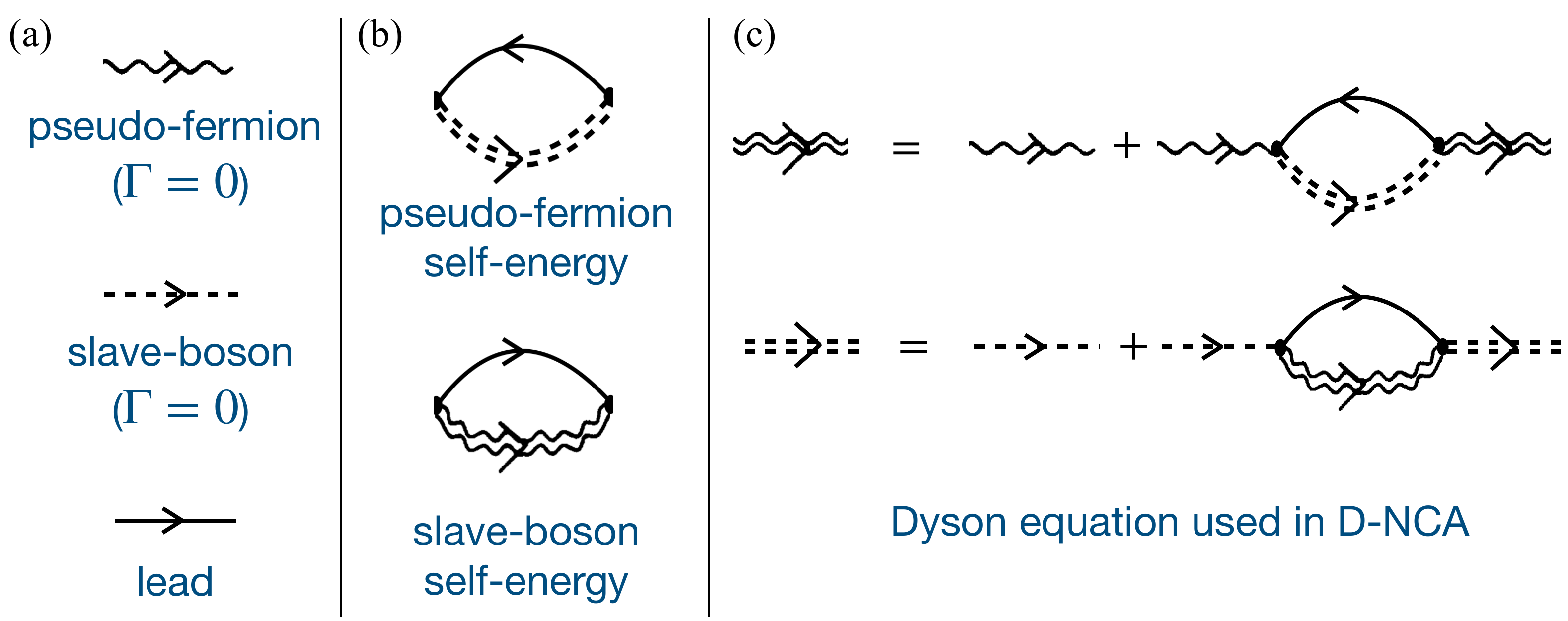}
		\caption{ 
			\label{fig:nca_diag}(a)Diagrammatic representation of the different propagators in presence of MBA. The  slave-boson and pseudo-fermion  propagators are the projected Green functions defined in Eq.~\eqref{eq:boson_gfn} and Eq.~\eqref{eq:fermion_gfn}. The single line denotes a MBA-modified bare propagator $(\Gamma=0)$ and  double line denotes exact propagator after including hybridization due to $\Gamma$. (b) Self-energy diagrams for the pseudo-fermion and slave-boson under D-NCA. (c) The diagrammatic representation of the Dyson equations for the pseudo-fermion  and slave-boson propagators under D-NCA.}
	\end{figure*}

	The self energy diagrams for the boson  ($\Pi(t)$) and fermion ($\Sigma_{f\sigma}(t)$) propagators under approximation (2) are shown in Fig.~\ref{fig:nca_diag}(b). Note that all the propagators (except for the lead) are the projected ones and so the self-energies are also the projected quantities. In particular, following Eq.~\eqref{eq:retarded_greater_relation} we have,
	\begin{equation}
		\Pi^>(\omega)=2i \,\text{Im}\left[\Pi^r(\omega)\right], \quad \Sigma_{f\sigma}^>(\omega)=2i \, \text{Im}\left[\Sigma_{f\sigma}^r(\omega)\right].
	\end{equation}
	
	We define the hybridization functions to the lead,
	\begin{eqnarray}
		\Delta_{\sigma}^>(\omega) &=& -2\pi i \gamma^2(1-f(\omega))\sum_k \delta(\omega-\epsilon_k)=-i\Gamma \frac{D^2}{\omega^2+D^2}(1-f(\omega)),\\
		\Delta_\sigma^<(\omega)&=&2\pi i \gamma^2 f(\omega)\sum_k \delta(\omega-\epsilon_k)=i\Gamma \frac{D^2}{\omega^2+D^2}f(\omega).
	\end{eqnarray}
	Here, we took the density of states (DoS) of the leads  to be a Lorentzian with width $D$ and consequently the line width (hybridization) function for the QD is given by
	\begin{equation}
		\Delta(\omega)=2\pi  \sum_k \gamma_k^2 \,\delta(\omega-\epsilon_k)=\Gamma \frac{D^2}{\omega^2+D^2}.
	\end{equation}
	Throughout we will work in the regime where $\epsilon_d, \Gamma, T \ll D$.

	The self-energy diagrams shown in Fig.~\ref{fig:nca_diag}(b) can be expressed as
	\begin{eqnarray}
		\Pi^{\overset{<}{>}}(\omega)&=&\frac{-i}{2\pi}\sum_\sigma\int_{-\infty}^{\infty}d\omega'\, \Delta_\sigma^{\overset{>}{<}}(\omega'-\omega)G_{f\sigma}^{\overset{<}{>}}(\omega'),\\
		\Sigma_{f\sigma}^{\overset{<}{>}}(\omega)&=&\frac{i}{2\pi}\int_{-\infty}^{\infty}d\omega'\, \Delta_\sigma^{\overset{<}{>}}(\omega-\omega')D^{\overset{<}{>}}(\omega').
	\end{eqnarray}
	The expression for the retarded self-energies follows from the relation between greater and retarded projected functions,
	\begin{eqnarray}
		\Pi^r(\omega)&=&\frac{i}{2\pi}\int_{-\infty}^{\infty}d\omega' \frac{\Pi^>(\omega')}{\omega-\omega'+i\eta},\\
		\Sigma_{f\sigma}^r(\omega)&=&\frac{i}{2\pi}\int_{-\infty}^{\infty}d\omega' \frac{\Sigma_{f\sigma}^>(\omega')}{\omega-\omega'+i\eta}.
	\end{eqnarray}
	The above form is not ideal for a numerical calculation. Fortunately, for the choice of Lorentzian DoS in the lead, a partial analytical integration is possible. To see this let us look at $\Pi^r(\omega)$ and use the form of $\Pi^>(\omega)$ to obtain
	\begin{eqnarray}
		\Pi^r(\omega) &=&\frac{1}{4\pi^2}\int_{-\infty}^{\infty} d\omega'\, \frac{1}{\omega-\omega'+i\eta}\int_{-\infty}^{\infty}d\omega''\left(\sum_{\sigma}\Delta_\sigma^<(\omega''-\omega')G_{f\sigma}^>(\omega'')\right)\\
		&=& \frac{i\Gamma D^2}{4\pi^2}\int_{-\infty}^{\infty}d\omega'' \, G_{f\sigma}^>(\omega'') \int_{-\infty}^\infty d\omega' \frac{f(\omega''-\omega')}{(\omega''-\omega')^2+D^2}\frac{1}{\omega-\omega'+i\eta},
	\end{eqnarray}
	with an analogous expression for $D^r(\omega)$. The final integral over $\omega'$ can be done analytically, yielding
	\begin{eqnarray}
		\Pi^r(\omega)&=&\sum_\sigma \int_{-\infty}^{\infty} d\omega'\, K(\omega'-\omega) G_{f\sigma}^>(\omega'),\\
		\Sigma_{f\sigma}^r(\omega)&=& \int_{-\infty}^{\infty} d\omega'\, K(\omega'-\omega) D^>(\omega').
	\end{eqnarray}
	The kernel $K(\omega)$ is given by
	\begin{equation}
		K(\omega)=\frac{\Gamma}{4\pi^2}\frac{D^2}{\omega^2+D^2}\left[\pi f(\omega)+ i \,\text{Re}\left[\psi\left(\frac{1}{2}-\frac{i\beta\omega}{2\pi}\right)-\psi\left(\frac{1}{2}+\frac{\beta D}{2\pi}\right)\right]-\frac{i\omega}{D}\left[\frac{\pi}{2}+\text{Im}\left[\psi\left(\frac{1}{2}+\frac{\beta D}{2\pi}\right)\right]\right]\right],
	\end{equation}
	where $\psi(z)$ is the digamma function.
	
	The Dyson equations for the projected Green functions are shown diagrammatically in Fig.~\ref{fig:nca_diag}(c). The solution of the Dyson equations for the projected retarded Green function is
	\begin{equation}
		\label{eq:dyson_retarded}
		D^r(\omega)=\frac{1}{\omega - \Pi^r(\omega)}, \quad
		G_{f\sigma}^r(\omega)=\frac{1}{(G_{f\sigma,0}^r(\omega;\epsilon_d))^{-1}-\Sigma_{f\sigma}^r(\omega)}.	
	\end{equation}
	The projected greater Green functions then are obtained using Eq.~\eqref{eq:retarded_greater_relation}. For the projected lesser Green functions, the solution of the Dyson equations is
	\begin{eqnarray}
		D^{<}(\omega)&=&\left[\left(1+D^r(\omega)\Pi^r(\omega)\right)D_0^<(
		\omega)\left(1+\Pi^a(\omega)D^a(\omega)\right)\right]+D^r(\omega)\Pi^{<}(\omega)D^a(\omega)\label{eq:dyson_d_less}, \\
		G_{f\sigma}^{<}(\omega)&=&\left[\left(1+G_{f\sigma}^r(\omega)\Sigma_{f\sigma}^r(\omega)\right)G_{f\sigma,0}^<(
		\omega)\left(1+\Sigma_{f\sigma}^a(\omega)G_{f\sigma}^a(\omega)\right)\right]+G_{f\sigma}^r(\omega)\Sigma_{f\sigma}^{<}(\omega)G_{f\sigma}^a(\omega)\label{eq:dyson_g_less}.
	\end{eqnarray}
	
	Note that the Dyson equations for the projected Green functions are obtained from the Dyson equations for the usual (grand-canonical) Green functions by substituting the Green function lines with the projected ones. The power counting consistency of this procedure is easy to see. The projected retarded function is $O(1)$ and it can be seen from Eq.~\eqref{eq:dyson_retarded} that the Dyson equation solution only involves retarded quantities and thereby satisfies the power counting consistency. The projected lesser function is $O(e^{-i\beta \mu})$ and it can be seen from Eq.~\eqref{eq:dyson_d_less} and \eqref{eq:dyson_g_less} that the Dyson equation solution only involves one lesser quantity in each term and thereby satisfies the power counting consistency. 
	
	In the standard NCA treatment in the absence of MBA, the bare lesser Green functions are delta-functions and this leads to the vanishing of the term in square brackets in Eq.~\eqref{eq:dyson_d_less} and \eqref{eq:dyson_g_less}.  In the presence of MBA in the D-NCA treatment, the term in the square bracket  vanish for the  boson Green function in Eq.~\eqref{eq:dyson_d_less} since the bare lesser function for the boson is a delta-function; but the term inside the square bracket for the pseudo-fermion Green function in Eq.~\eqref{eq:dyson_g_less} does not vanish.
	
	Note that the exact propagators are expressed in terms of the self-energies, which in turn are expressed in terms of the exact propagators. We obtain a self-consistent solution  numerically. For clarity, we now detail the sequence of steps followed to obtain the self-consistent solution. 
	
	\subsection{Procedure for self-consistent steady state solution in D-NCA}
	The procedure consists of two separated self-consistent loops. One for the retarded/greater functions and another for the lesser functions. The self-consistent loop for the retarded/greater functions does not involve the lesser functions. So this loop is first numerically solved. The self-consistent loop for the lesser functions involve the greater functions. We use the solutions obtained from the greater loop for this. 
	\subsubsection{Self-consistent loop for retarded/greater functions}
	\begin{enumerate}
		\item Start with a guess for $D^>(\omega)$. We choose, $D^>(\omega)=\frac{1}{\omega+i\Gamma}$
		\item $\Sigma_{f\sigma}^r(\omega)=\int_{-\infty}^{\infty}d\omega' \, K(\omega'-\omega)D^>(\omega')$ 
		\item $G_{f\sigma}^r(\omega)=\frac{1}{(G_{f\sigma,0}^r(\omega;\epsilon_d))^{-1}-\Sigma_{f\sigma}^r(\omega)}$
		\item $G_{f\sigma}^>(\omega)=2i\,\text{Im} (G_{f\sigma}^r(\omega))$
		\item $\Pi^r(\omega)=\sum_\sigma \int_{-\infty}^{\infty}d\omega'\, K(\omega'-\omega)G_{f\sigma}^>(\omega')$
		\item $D^r(\omega)=\frac{1}{\omega-\Pi^r(\omega)}$
		\item $D^>(\omega)=2i\, \text{Im}(D^r(\omega))$
		\item Go back to 2 until convergence for $G^>_{f\sigma}(\omega)$ and $D^>(\omega)$ is obtained.
	\end{enumerate}
	\subsubsection{Self-consistent loop for lesser functions}
	\begin{enumerate}
		\item Start with a guess for $D^<(\omega)$. We choose, $D^<(\omega)=\frac{1}{\omega+i\Gamma}$
		\item $\Sigma_{f\sigma}^<(\omega)=\frac{i}{2\pi}\int_{-\infty}^{\infty}d\omega'\, \Delta_{\sigma}^<(\omega-\omega')D^<(\omega')$
		\item $G_{f\sigma}^{<}(\omega)=\left[\left(1+G_{f\sigma}^r(\omega)\Sigma_{f\sigma}^r(\omega)\right)G_{f\sigma,0}^<(
		\omega)\left(1+\Sigma_{f\sigma}^a(\omega)G_{f\sigma}^a(\omega)\right)\right]+G_{f\sigma}^r(\omega)\Sigma_{f\sigma}^{<}(\omega)G_{f\sigma}^a(\omega)$
		\item $\Pi^<(\omega)=\frac{-i}{2\pi}\sum_{\sigma}\int_{-\infty}^{\infty}d\omega' \Delta_\sigma^>(\omega'-\omega)G_{f\sigma}^<(\omega')$
		\item $D^{<}(\omega)=D^r(\omega)\Pi^{<}(\omega)D^a(\omega)$
		\item Go back to 2 until convergence for $G_{f\sigma}^<(\omega)$ and $D^<(\omega)$ is obtained.
	\end{enumerate}
	
	\subsubsection{Consistency checks}
	The following consistency checks are used on the numerically obtained self-consistent solution to make sure proper convergence is obtained: 
	\begin{eqnarray}
		\frac{-1}{\pi}\int_{-\infty}^\infty	d\omega\, \text{Im}[D^r(\omega)]&=&1, \\
		\frac{-1}{\pi}\int_{-\infty}^\infty	d\omega\, \text{Im}[G_{f\sigma}^r(\omega)]&=&1, \\
		\int_{-\infty}^{\infty} d\omega \rho_\sigma(\omega)&=& 1-\sum_{\sigma' \ne \sigma}\left< n_{\sigma'}\right> .
	\end{eqnarray}
	These consistency checks are the same as in standard NCA, and we refer the reader to Ref.~\cite{wingreen1994anderson} for further details.
	
	\subsection{ \label{sec:physical_observables}Physical observables and correlators of the original fermions}
	We list below how the different physical observables and correlators of the original fermions that we are interested in are obtained using the D-NCA solution:
	\begin{eqnarray}
		\left<n_\sigma\right>&=&\frac{Z_{Q=0}}{Z_{Q=1}}\left[\frac{-i}{2\pi}\int_{-\infty}^{\infty}d\omega\, G_{f\sigma}^<(\omega)\right], \quad \text{with}, \\
		\frac{Z_{Q=1}}{Z_{Q=0}}&=&\frac{i}{2\pi}\int_{-\infty}^{\infty}d\omega \left[D^<(\omega)-\sum_\sigma G_{f\sigma}^<(\omega)\right] \label{eq:z_ratio},\\
		G_{\sigma}^>(t)&=&\frac{iZ_{Q=0}}{Z_{Q=1}} D^<(-t)G_{f\sigma}^>(t)  \implies G_\sigma^>(\omega)=\frac{i}{2\pi}\frac{Z_{Q=0}}{Z_{Q=1}}\int_{-\infty}^\infty d\omega'\, D^<(\omega')G_{f\sigma}^>(\omega+\omega')\label{eq:G_great_rel},\\
		G_{\sigma}^<(t)&=& \frac{iZ_{Q=0}}{Z_{Q=1}} D^>(-t) G_{f\sigma}^<(t)  \implies G_\sigma^<(\omega)=\frac{i}{2\pi}\frac{Z_{Q=0}}{Z_{Q=1}}\int_{-\infty}^\infty d\omega'\, D^>(\omega')G_{f\sigma}^<(\omega+\omega')\label{eq:G_less_rel}, \\
		G_{\sigma}^r(t)&=&\theta(t)\left[G_\sigma^>(t)-G_\sigma^<(t)\right],\\
		\rho_\sigma(\omega)&=& \frac{1}{4\pi^2}\frac{Z_{Q=0}}{Z_{Q=1}} \int_{-\infty}^\infty d\omega'\, \left[D^>(\omega')G_{f\sigma}^<(\omega+\omega')-D^<(\omega')G_{f\sigma}^>(\omega+\omega')\right].
	\end{eqnarray}
	
	\subsection{Current conservation in D-NCA}
	Current conservation is  an important physical requirement. 
	Current conservation for the setup that we are considering means the average electric current between the impurity QD and the lead vanishes in the steady state. Note that if more than one leads are coupled to the QD then the current conservation condition means that the sum of the currents to the leads vanishes. The proof that D-NCA is a current conserving approximation is identical for the cases of single and multiple leads. We discuss the single lead case for simplicity.
	
	The defining expression for the electric current from the impurity QD to the lead is ~\cite{wingreen1994anderson},
	\begin{equation}
		J=\frac{1}{2\pi}\sum_\sigma\int_{-\infty}^\infty d\omega\, \left[\Delta_\sigma^<(\omega)G_\sigma^>(\omega)-\Delta_\sigma^>(\omega)G_\sigma^<(\omega)\right].
	\end{equation}
	Plugging in the expression of the electron Green functions in terms of the slave boson and fermion Green functions from Eqns.~\eqref{eq:G_great_rel} and \eqref{eq:G_less_rel}, we obtain
	\begin{equation}
		J \propto \sum_\sigma \int_{-\infty}^{\infty} d\omega'\, \int_{-\infty}^\infty d\omega'\left[\Delta_\sigma^<(\omega)D^<(\omega')G_{f\sigma}^>(\omega+\omega')-\Delta_\sigma^>(\omega)D^>(\omega')G_{f\sigma}^<(\omega+\omega') \right].
	\end{equation}	
	We can rewrite the above using the expression for the boson self energies as
	\begin{equation}
		J \propto \int_{\-\infty}^{\infty} d\omega' \left[D^>(\omega')\Pi^<(\omega')-D^<(\omega')\Pi^>(\omega')\right].
	\end{equation}
	Since the boson Green functions obey
	\begin{equation}
		D^{\overset{>}{<}}(\omega)=D^r(\omega) \Pi^{\overset{>}{<}}(\omega)D^a(\omega),
	\end{equation}
	we see that $J=0$, hence  D-NCA is a current conserving approximation.
	\subsection{\label{sec:dnca_independence} Independence of the D-NCA steady state solution on initial occupation probabilities}
	
	The hybridization expansion assumes that the tunneling is turned on abruptly at some point in time and then the system evolves to a steady state under the full Hamiltonian. Physically one expects the steady state solution to be independent of the initial occupation probabilities, $P_\sigma$ at $\Gamma=0$. But in the self-consistent loop for the projected lesser Green functions, $G_{0,f\sigma}^<(\omega)$ enters, which from Eq.~\eqref{eq:bare_fermion} carries the occupation probability $P_\sigma$. So does it mean that the D-NCA steady state solution depends on  $P_\sigma$? The answer is no. This follows from the fact that the overall normalization of the projected lesser Green function is arbitrary. In fact, in all the physical quantities related to the original fermions that are defined in the section~\ref{sec:physical_observables}, the projected lesser functions are always accompanied by the factor $Z_{Q=0}/Z_{Q=1}$, and then it follows from Eq.~\eqref{eq:z_ratio} that the overall normalization of the projected lesser functions dropout. Next, from the equations in the self-consistent loop for the lesser functions, it can be seen that if $P_\sigma \to c \, P_\sigma$ in the bare projected lesser pseudo-fermion Green function, with $c$ being a non-zero complex scalar, the self-consistent solutions simply scales by the same factor, i.e. $D^<(\omega),G_{f\sigma}^<(\omega) \to c \,D^<(\omega), c \, G_{f\sigma}^<(\omega)$. Since the overall normalization of the lesser functions dropout for physical quantities, it then follows that the D-NCA steady state solution is independent of $P_\sigma$. Using this fact one can simply take the projected bare pseudo-fermion lesser Green function as,
	\begin{equation}
		G_{f\sigma,0}^<(\omega)=iA^-(\omega-\epsilon_d).
	\end{equation}
	
	\subsection{Verification that D-NCA is exact in the  limit $\Gamma\to 0$}
	
	We show that the D-NCA steady state solution in the limit $\Gamma \to 0$ 
	correctly gives the exact electron Green function  with the correct occupation probabilities, as discussed in Section~\ref{sec:exact_weak_tun}. In the  limit $\Gamma \to 0$, the pseudo-fermion functions simply take the bare form and so,
	\begin{equation}
		G_{f\sigma}^>(\omega) = -i A^+(\omega-\epsilon_d), \quad 
		G_{f\sigma}^<(\omega) = i A^-(\omega-\epsilon_d).
	\end{equation}
	The boson  self energies are,
	\begin{eqnarray}
		\Pi^>(\omega)&=&\frac{\Gamma}{2\pi} \sum_\sigma \int d\omega' \, f(\omega'-\omega)G_{f\sigma}^>(\omega') = \frac{-i}{\pi} \Gamma_{in}(\epsilon_d-\omega),\\
		\Pi^<(\omega) &=& \frac{-\Gamma}{2\pi} \sum_\sigma \int d\omega' \, (1-f(\omega'-\omega))G_{f\sigma}^<(\omega') = \frac{-i}{\pi} \Gamma_{out}(\epsilon_d-\omega).
	\end{eqnarray}
	In the limit, $\Gamma \to 0$, we only need the imaginary part of the boson retarded self energy, which is easily obtained from the relation $\Pi^>(\omega)=2i\text{Im}(\Pi^r(\omega))$. Thus,
	\begin{equation}
		\Pi^r(\omega)=  \frac{-i}{2\pi} \Gamma_{in}(\epsilon_d-\omega),
	\end{equation}
	and then the retarded boson function becomes
	\begin{equation}
		D^r(\omega)=\frac{1}{\omega+\frac{i}{2\pi}\Gamma_{in}(\epsilon_d-\omega)}.
	\end{equation}
	The Greater  and lesser boson functions are then obtained as,
	\begin{eqnarray}
		D^>(\omega)&=&|D^r(\omega)|^2\Pi^>(\omega) \xrightarrow{\Gamma \to 0} -2\pi i \delta(\omega),\\
		D^<(\omega) &=& |D^r(\omega)|^2 \Pi^<(\omega) \xrightarrow{\Gamma \to 0} -2\pi i \frac{\Gamma_{out}(\epsilon_d)}{\Gamma_{in}(\epsilon_d)}\delta(\omega).
	\end{eqnarray}
	Plugging in the forms of the slave boson and fermion Green functions to Eq.~\eqref{eq:z_ratio}, we get,
	\begin{equation}
		\frac{Z_{Q=1}}{Z_{Q=0}}= \frac{2\Gamma_{in}(\epsilon_d)+\Gamma_{out}(\epsilon_d)}{\Gamma_{in}(\epsilon_d)}.
	\end{equation}
	Then from Eqs.~\eqref{eq:G_great_rel} and \eqref{eq:G_less_rel}, we see that D-NCA steady state solution gives the exact electron Green functions in the $\Gamma \to 0$ limit as discussed in section~\ref{sec:exact_weak_tun}.
	
	\subsection{Energy current}
	Following Ref.~\cite{meir1992landauer}, the electric current that flows between the QD and its lead can be expressed as
	\begin{equation}
		J=e\Gamma \sum_\sigma\int d\omega\, \left[\left(G_{\sigma}^r(\omega)-G_{\sigma}^a(\omega)\right)f(\omega) + G_{\sigma}^<(\omega)\right].
	\end{equation}
	Using Eq.~\eqref{eq:dos_def}  and Eq.~\eqref{eq:h_def} 
	we can rewrite the expression as
	\begin{equation}
		J=2e\Gamma \int d\omega\, \nu(\omega) (h(\omega)-f(\omega)).
	\end{equation}
	Since $\omega=0$ corresponds to the Fermi energy of the lead, the energy current density that flows in to the lead at an energy $\omega$ is simply obtained by multiplying the corresponding particle current density with $\omega$. Thus the total energy current that flows into the lead can be expressed as
	\begin{equation}
		J_{\rm{en}}=2\Gamma \int d\omega\, \omega\, \nu(\omega) (h(\omega)-f(\omega)).
	\end{equation}
	By the steady state condition, this is the energy current that flows into the QD from the detector.